\let\oldvec\vec
\documentclass[runningheads]{llncs}
\let\vec\oldvec

\usepackage{graphicx} % Required for including images
\usepackage{booktabs} % Top and bottom rules for table
\usepackage{wrapfig} % Allows in-line images
\usepackage{cite}
\usepackage[utf8]{inputenc} % allow utf-8 input
\usepackage[T1]{fontenc}    % use 8-bit T1 fonts
\usepackage{hyperref}       % hyperlinks
\usepackage{url}            % simple URL typesetting
\usepackage{booktabs}       % professional-quality tables
\usepackage{nicefrac}       % compact symbols for 1/2, etc.
\usepackage{microtype}      % microtypography
\usepackage{lipsum}
\graphicspath{ {./images/} }
\usepackage{subfigure}
\usepackage{amsfonts, amsmath, amssymb} % For math fonts, symbols and environments
\usepackage{booktabs} % Top and bottom rules for table
\usepackage{wrapfig} % Allows in-line images
\usepackage{sidecap}
\sidecaptionvpos{figure}{t}
\usepackage{indentfirst}
\usepackage[vlined,ruled]{algorithm2e}

\usepackage{array}
\newcolumntype{L}[1]{>{\raggedright\let\newline\\\arraybackslash\hspace{0pt}}p{#1}}
\newcolumntype{C}[1]{>{\centering\let\newline\\\arraybackslash\hspace{0pt}}p{#1}}
\newcolumntype{R}[1]{>{\raggedleft\let\newline\\\arraybackslash\hspace{0pt}}p{#1}}

\usepackage{adjustbox}

\usepackage{textcomp}

\usepackage{soul}
\usepackage{color}

\usepackage{threeparttable}

\usepackage{enumitem}
\usepackage{multirow}
\setlist[description]{leftmargin=8pt,labelindent=0pt,itemsep=0pt}
\setlist[itemize]{itemsep=0pt,parsep=0pt}
\setlist[enumerate]{itemsep=0pt,parsep=0pt, topsep=0pt}

\title{Fundus2Angio: A Conditional GAN Architecture for Generating Fluorescein Angiography Images from Retinal Fundus Photography}
\titlerunning{Fundus2Angio}
\author{Sharif Amit Kamran\inst{1} \and 
Khondker Fariha Hossain\inst{2}\and  Alireza Tavakkoli\inst{1} \and Stewart Zuckerbrod\inst{3} \and Salah A. Baker\inst{4} \and Kenton M. Sanders\inst{4}}
\authorrunning{S.A. Kamran, et al.}
\institute{University of Nevada, Reno, NV 89557\\
\and
Deakin University, Melbourne, AUS\\
\and
Houston Eye Associates, Houston, TX 77801\\
\and
University of Nevada School of Medicine, Reno, NV 89557}

\begin{document}

\maketitle              % typeset the header of the contribution
\begin{abstract}
Carrying out clinical diagnosis of retinal vascular degeneration using Fluorescein Angiography (FA) is a time consuming process and can pose significant adverse effects on the patient. Angiography requires insertion of a dye that may cause severe adverse effects and can even be fatal. Currently, there are no non-invasive systems capable of generating Fluorescein Angiography images. However, retinal fundus photography is a non-invasive imaging technique that can be completed in a few seconds. In order to eliminate the need for FA, we propose a conditional generative adversarial network (GAN) to translate fundus images to FA images. The proposed GAN consists of a novel residual block capable of generating high quality FA images. These images are important tools in the differential diagnosis of retinal diseases without the need for invasive procedure with possible side effects. Our experiments show that the proposed architecture achieves a low FID score of 30.3 and outperforms other state-of-the-art generative networks. Furthermore, our proposed model achieves better qualitative results indistinguishable from real angiograms.
\end{abstract}
\section{Introduction}
%\vspace{-4mm}\section{Introduction}\vspace{-1mm}
\label{sec:intro}

For a long time Fluorescein Angiography (FA) combined with Retinal Funduscopy have been used for diagnosing retinal vascular and pigment epithelial-choroidal diseases \cite{mary2016retinal}. The process requires the injection of a fluorescent dye which appears in the optic vein within 8-12 seconds depending on the age and cardiovascular structure of the eye and stays up to 10 minutes \cite{mandava2004fluorescein}. Although generally considered safe, there have been reports of mild to severe complications due to allergic reactions to the dye \cite{brockow2014hypersensitivity}. Frequent side effects can range from nausea, vomiting, anaphylaxis, heart attack, to anaphylactic shock and death \cite{lira2007adverse}. In addition, leakage of fluorescein in intravaneous area is common. However, the concentration of fluorescein solutions don't have any direct impact on adverse effects mentioned above.\cite{yannuzzi1986fluorescein}.

Given the complications and the risks associated with this procedure, a non-invasive, affordable, and computationally effective procedure is quite imperative. The only current alternatives to flourecein angigraphy (FA) is carried out by Optical Coherence Tomography and basic image processing technique. These systems are generally quite expensive. Without a computationally effective and financially viable mechanism to generate reliable and reproducible flourecein angiograms, the only alternative is to utilize retina funduscopy for differential diagnosis. Although automated systems consisting of image processing and machine learning algorithms have been proposed for diagnosing underlying conditions and diseases from fundus images \cite{poplin2018prediction}, there has not been an effective effort to generate FA images from retina photographs. In this paper, we propose a novel conditional Generative Adversarial Network (GAN) called Fundus2Angio, capable of synthesizing fluorescein angiograms from retinal fundus images. The procedure is completely automated and does not require any human intervention. We use both qualitative and quantitative metrics for testing the proposed architecture. We compare the proposed architecture with other state-of-the-art conditional GANs~\cite{wang2018high,isola2017image,zhu2017unpaired}. Our model outperforms these networks in terms of quantitative measurement. For qualitative results, expert ophthalmologists were asked to distinguish fake angiograms from a random set of balanced real and fake angiograms over two trials. Results show that the angiograms generated by the proposed network are quite indistinguishable from real FA images.

\section{Literature Review}
%\vspace{-4mm}\section{Literature Review}\vspace{-1mm}

Generative adversarial networks have revolutionized many image manipulation tasks such as image editing \cite{zhu2016generative}, image styling \cite{chen2018sketchygan}, and image style transfer \cite{zhu2017unpaired,wang2018high}. Multi-resolution architectures are common practice in computer vision, while coupled architectures have the capability to combine fine and coarse information from images \cite{brown2003recognising}. 
Recently, techniques on Conditional \cite{huang2017stacked} and Unconditional GANs \cite{chen2017photographic} have explored the idea of combined-resolutions within the architecture for domain specific tasks. Inspired by this, we propose an architecture that extract features at different scales.

Some approaches also used multi-scale discriminators for style-transfer \cite{wang2018high}. However, they only attached discriminators with generator that deals with fine features while ignoring discriminators for coarse generator completely. In order to learn useful features at coarsest scale, separate multi-scale discriminators are necessary. Our proposed architecture employs this for both coarse and fine generators. 

\begin{figure}[htb]
    \centering
    \includegraphics[width=\columnwidth]{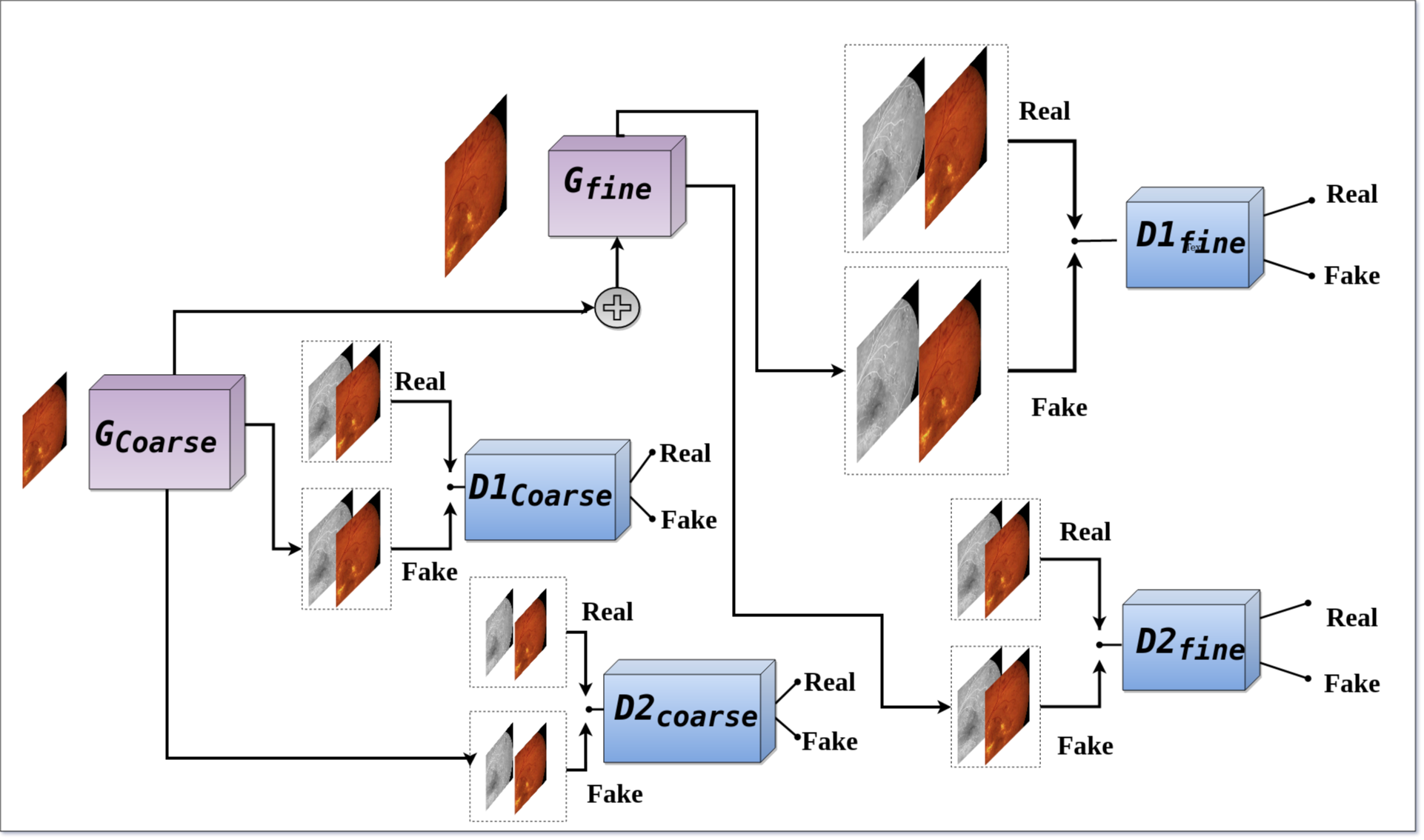}
    \caption{Proposed Generative Adversarial Network consisting of two Generators $G_{coarse}$, $G_{fine}$, and four discriminators $D1_{coarse}$, $D1_{fine}$, $D2_{fine}$, $D2_{coarse}$. The generators take Fundus image as input and outputs FA image. Whereas, the discriminators take both Fundus and FA images as input and outputs if the pairs are real or fake. }
    \label{fig1}
\end{figure}

For high quality image synthesis, a pyramid network with multiple pairs of discriminators and generators has also been proposed, termed SinGAN \cite{shaham2019singan}. Though it produces high quality synthesized images, the model works only on unpaired images. To add to this problem, each generator's input is the synthesized output produced by the previous generator. As a result, it can't be employed for pair-wise image training that satisfies a condition. To alleviate from this problem, a connection needs to be established that can propagate feature from coarse to fine generator. In this paper, we propose such an architecture that has a feature appending mechanism between the coarse and fine generators, making it a two level pyramid network with multi-scale discriminators as illustrated in Fig.~\ref{fig1}. 

\section{The Proposed Methodology}
This paper introduces a new conditional generative adversarial network (GAN) comprising of a novel residual block for producing realistic FA from retinal fundus images. First, we introduce the residual block in section \ref{subsec:residualblock}. We then delve into the proposed conditional GAN encompassing of fine and coarse generators and four multi-scale discriminators in sections \ref{subsec:generators} and \ref{subsec:discriminators}. Lastly, in section \ref{subsec:objective}, we discuss the objective function and loss weight distributions for each of the architectures that form the proposed architecture.

\subsection{Novel Residual Block}
%\vspace{-2mm}\subsection{Novel Residual Block}\vspace{-1mm}
\label{subsec:residualblock}

Recently, residual blocks have become the norm for implementing many image classification, detection and segmentation architectures \cite{he2016identity}. Generative architectures have employed these blocks in interesting applications ranging from image-to-image translation to super-resolution \cite{johnson2016perceptual,wang2018high}. In its atomic form, a residual unit consists of two consecutive convolution layers. The output of the second layers is added to the input, allowing for deeper networks. Computationally, regular convolution layers are expensive compared to a newer convolution variant, called separable convolution \cite{chollet2017xception}. Separable convolution performs a depth-wise convolution followed by a point-wise convolution. This, in turn helps to extract and retain depth and spatial information through the network. It has been shown that interspersing convolutional layers allows for more efficient and accurate networks~\cite{opticnet19}. We incorporate this idea to design a novel residual block to retain both depth and spatial information, decrease computational complexity and ensure efficient memory usage, as shown in Table.~\ref{table1}.

\begin{table}[htb]
\caption{Comparison between Original and Proposed Residual Block}
    \label{table1}
\centering
\begin{tabular}{l|c|c|c} 
\hline
\small Residual Block & \small Equation & \small Activation & \small No. of Parameters$^{1}$\\
\hline
\small Original
& \small $\big[R_{i}  \circledast F_{Conv} \circledast F_{Conv} \big] + R_{i}$ & \small ReLU (Pre) \cite{he2016identity}& \small 18,688 \\ 
\small Proposed  & \small $\big[R_{i}  \circledast F_{Conv} \circledast F_{SepConv} \big] + R_{i}$ &  \small Leaky-ReLU (Post) & \small 10,784\\ 
\hline
\end{tabular}
\footnotesize
\\$^1$ $F_{Conv}$ and $F_{SepConv}$ has kernel size $K=3$, stride $S=1$, padding $P=0$ and No. of channel $C=32$.
\end{table}

\begin{figure}[htb]
    \centering
    \includegraphics[width=\columnwidth]{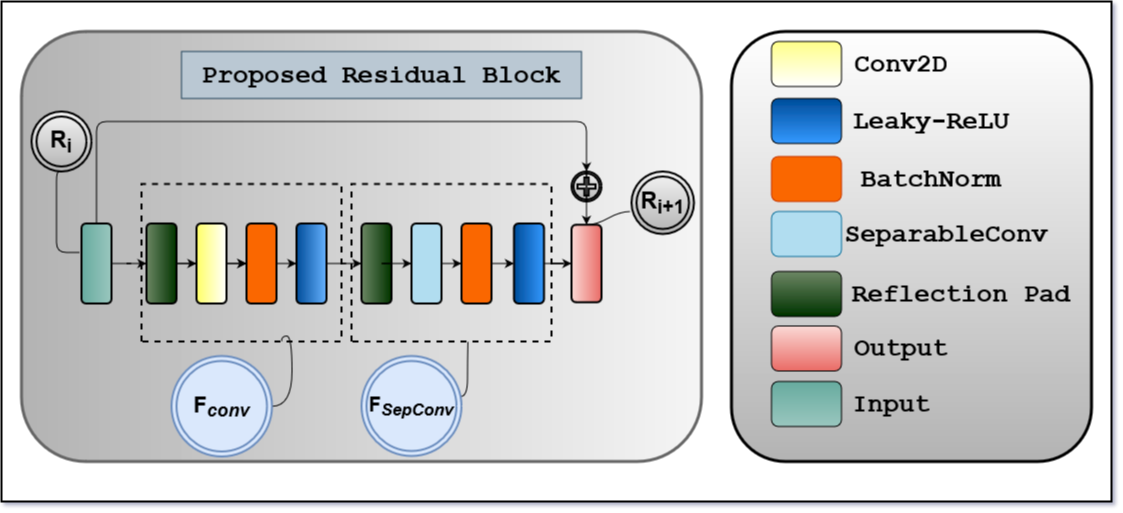}
    \caption{Proposed Residual Block consisting of two residual units $F_{conv}$ and $F_{SepConv}$. First one cosists of Reflection padding, Convolution, Batch-Normalization and Leaky-ReLU layers. The second one has some layers except has Separable Convolution instead of Vanilla Convolution. $R_i$ and $R_{i+1}$ signifies input and output of the residual block} 
    \label{fig2}
\end{figure}

As illustrated in Fig.~\ref{fig2}, we replace the last convolution operation with a separable convolution. We also use Batch-normalization and Leaky-ReLU as post activation mechanism after both convolution and separable Convolution layers. For better results, we incorporate reflection padding as opposed to zero-padding before each convolution operation. The entire operation can be formulated as shown in Eq.~\ref{eq1}:
\begin{equation}
\begin{split}
R_{i+1} &=  \big[R_{i}  \circledast F_{Conv} \circledast F_{SepConv} \big] + R_{i} 
\\ &=   F(R_{i}) + R_{i}
\label{eq1}
\end{split}
\end{equation}

Here, $\circledast$ refers to convolution operation while $F_{conv}$ and $F_{SepConv}$ signify the back-to-back convolution and separable convolution operations. By exploiting convolution and separable convolution layer with Leaky-ReLU, we ensure that two distinct feature maps (spatial \& depth information) can be combined to generate fine fluorescein angiograms. 

\begin{figure}[htb]
    \centering
    \includegraphics[width=\columnwidth]{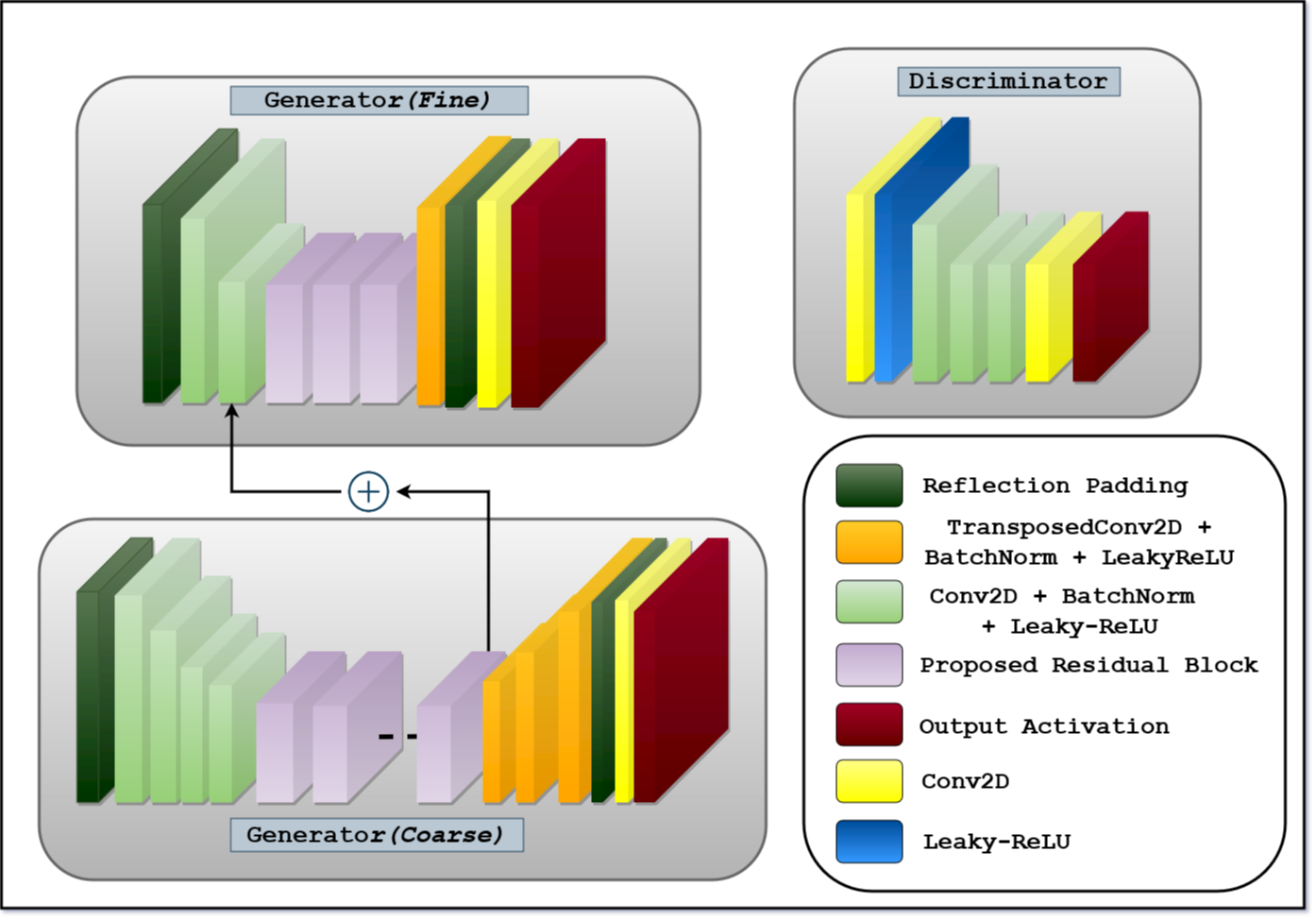}
    \caption{The backbone for $G_{fine}$, $G_{coarse}$ Generators and $D1$, $D2$  Discriminator Architectures. The $G_{fine}$ consists of two encoding blocks (light green), followed by three residual blocks (purple) and one decoding block (orange).The $G_{coarse}$ consists of four encoding blocks (light green), followed by nine residual blocks (purple) and three decoding blocks (orange). The discriminator consists of three encoding block (light green). Both the generators have intermediate Conv2D, Reflection padding layer and TanH as output activation function. Whereas, discriminators have intermediate Conv2D, Leaky-ReLU layers and Sigmoid as output activation.}
    \label{fig3}
\end{figure}

\subsection{Coarse and Fine Generators}\label{subsec:generators}

Using a coarse-to-fine generator for both conditional and unconditional GANs results in very high quality image generation, as observed in recent architectures, such as pix2pixHD \cite{wang2018high} and SinGan \cite{shaham2019singan}. Inspired by this idea, we use two generators ($G_{fine}$ and $G_{coarse}$) in the proposed network, as illustrated in Fig.~\ref{fig3}. The generator $G_{fine}$ synthesizes fine angiograms from fundus images by learning local information, including retinal venules, arterioles, hemorrhages, exudates and microaneurysms. On the other hand, the generator $G_{coarse}$ tries to extract and preserve global information, such as the structures of the macula, optic disc, color, contrast and brightness, while producing coarse angiograms. 

The generator $G_{fine}$  takes input images of size $512\times 512$ and produces output images with the same resolution. Similarly, the generator $G_{coarse}$ network takes an image with half the size ($256\times 256$) and  outputs an image of the same size as the input. In addition, the $G_{coarse}$ outputs a feature vector of the size $256\times 256 \times 64$ that is eventually added with one of the intermediate layers of $G_{fine}$. These hybrid generators are quite powerful for sharing local and global information between multiple architectures as seen in \cite{johnson2016perceptual,shaham2019singan,wang2018high}. Both generators use convolution layers for downsampling and transposed convolution layers for upsampling. It should be noted that $G_{coarse}$ is downsampled twice ($\times 2$) before being upsampled twice again with transposed convolution. In both the generators, the proposed residual blocks are used after the last downsampling operation and before the first upsampling operations as illustrated in Fig.~\ref{fig3}. On the other hand, in $G_{fine}$, downsampling takes place once with necessary convolution layer, followed by adding the feature vector, repetition of residual blocks and then upsampling to get fine angiography image. All convolution and transposed convolution operation are followed by Batch-Normalization and Leaky-ReLU activations. To train these generators, we start with $G_{coarse}$ by batch-training it on random samples once and then we train the $G_{fine}$ once with a new set of random samples. During this time, the discriminator's weights are frozen, so that they are not trainable. Lastly, we jointly fine-tune  all the discriminator and generators together to train the GAN.

\subsection{Multi-scale PatchGAN as Discriminator}\label{subsec:discriminators}

For synthesizing fluorescein angiography images, GAN discriminators need to adapt to coarse and fine  generated images for distinguishing between real and fake images. To alleviate this problem, we either need a deeper architecture or, a kernel with wider receptive field. Both these solutions result in over fitting and increase the number of parameters. Additionally, a large amount of processing power will be required for computing all the parameters. To address this issue, we exploit the idea of using two Markovian discriminators, first introduced in a technique called PatchGAN \cite{li2016precomputed}. This technique takes input from different scales as previously seen in \cite{wang2018high,shaham2019singan}. 

We use four discriminators that have a similar network structure but operate at different image scales. Particularly, we downsample the real and generated angiograms by a factor of $2$ using the Lanczos sampling to create an image pyramid of three scales (original and $2\times$downsampled and $4\times$downsampled). We group the four discriminators into two, $D_{fine}=[D1_{fine},D2_{fine}]$ and $D_{coarse}=[D1_{coarse},D2_{coarse}]$ as seen in Fig.~\ref{fig1}. The discriminators are then trained to distinguish between real and generated angiography images at the three distinct resolutions respectively. 

The outputs of the PatchGAN for $D_{fine}$ are $64\times64$ and $32\times32$ and for $D_{coarse}$ are $32\times32$ and $16\times16$. With the given discriminators, the loss function can be formulated as given in Eq.~\ref{eq2}. It's a multi-task problem of maximizing the loss of the discriminators while minimizing the loss of the generators. 
\begin{equation}
    \min \limits_{G_{fine},G_{coarse}} \max \limits_{D_{fine},D_{coarse}}  \mathcal{L}_{cGAN}(G_{fine},G_{coarse}, D_{fine},D_{coarse})
    \label{eq2}
\end{equation}

Despite discriminators having similar network structure, the one that learns feature at a lower resolution has the wider receptive field.  It tries to extract and retain more global features such as macula, optic disc, color and brightness etc to generate better coarse images. In contrast,  the  discriminator that learns feature at original resolution dictates the generator to produce fine features such as retinal veins and arteries, exudates etc. By doing this we combine feature information of global and local scale while training the generators independently with their paired multi-scale discriminators.

\subsection{Weighted Object Function and Adversarial Loss}\label{subsec:objective}

We use LSGAN \cite{mao2017least} for calculating the loss and training our conditional GAN. The objective function for our conditional GAN is given in Eq.~\ref{eq3}. 

\begin{equation}
    \mathcal{L}_{cGAN}(G,D) =  \mathbb{E}_{x,y} \big[\ (D(x,y) -1)^2 \big]\ +  \mathbb{E}_{x} \big[\ (D(x,G(x)+1))^2 \big]\
\label{eq3}
\end{equation}
where the discriminators are first trained on the real fundus, $x$ and real angiography image, $y$ and then trained on the the real fundus, $x$ and fake angiography image, $G(x)$. We start with training the discriminators $D_{fine}$ and $D_{coarse}$ for couple of iterations on random batches of images. Next, we train the $G_{coarse}$ while keeping the weights of the discriminators frozen. Following that, we train the the $G_{fine}$ on a batch of random samples in a similar fashion. We use Mean-Squared-Error (MSE) for calculating the individual loss of the generators as shown in Eq.~\ref{eq4}.

\begin{equation}
    \mathcal{L}_{L2}(G) = \mathbb{E}_{x,y} \Vert G(x) - y \Vert^2
    \label{eq4}
\end{equation}

where, $\mathcal{L}_{L2}$ is the reconstruction loss for a real angiogram, $y$, given a generated angiogram, $G(x)$. We use this loss for both $G_{fine}$ and $G_{coarse}$ so that the model can generate high quality angiograms of different scales. Previous techniques have also exploited this idea of combining basic GAN objective with a MSE loss \cite{pathak2016context}. From Eq.~\ref{eq3} and \ref{eq4} we can formulate our final objective function as given in Eq.~\ref{eq5}.

\begin{equation}
\begin{split}
    \min \limits_{G_{fine},G_{coarse}} \max \limits_{D_{fine},D_{coarse}}  \mathcal{L}_{cGAN}(G_{fine},G_{coarse}, D_{fine},D_{coarse}) \\+ \lambda\big[\ \mathcal{L}_{L2}(G_{fine}) + \mathcal{L}_{L2}(G_{coarse})\big]\
    \label{eq5}
\end{split}
\end{equation}

Here, $\lambda$ dictates either to prioritize the discriminators or the generators. For our architecture, more weight is given to the reconstruction loss of the generators and thus we pick a large $\lambda$ value.

\section{Experiments}\label{sec:Results}

In the following section, different experimentation and evaluation is provided for our proposed architecture. First we elaborate about the data preparation and pre-prossessing scheme in Sec.~\ref{subsec:dataset}. We then define our hyper-parameter settings in Sec.~\ref{subsec:hyper}. Following that, different architectures are compared based on some quantitative and qualitative evaluation metrics in Sec.~\ref{subsec:quant}. Lastly, and Sec.~\ref{subsec:qual},

\subsection{Dataset}\label{subsec:dataset}
For training, we use the funuds and angiography data-set provided by Hajeb et al. \cite{hajeb2012diabetic}. The data-set consists of 30 pairs of diabetic retinopathy and 29 pairs normal of angiography and fundus images from 59 patients. Because, not all of the pairs are perfectly aligned, we select 17 pairs for our experiment based on alignment. The images are either perfectly aligned or nearly aligned. The resolution for fundus and angiograms are as follows $576\times720$. Fundus photographs are in RGB format, whereas angiograms are in Gray-scale format. Due to shortage of data, we take 50 random crops of size $512\times512$ from each images for training our model. So, the total number of training sample is 850 ($17\times50$). 

\subsection{Hyper-parameter tuning}
\label{subsec:hyper}
LSGAN \cite{mao2017least} was found to be effective for generating desired synthetic images for our tasks. We picked $ \lambda =10$ (Eq.~\ref{eq5}). For optimizer, we used Adam with learning rate $\alpha=0.0002$, $\beta_1=0.5$ and $\beta_2=0.999$. We train with mini-batches with batch size, $b=4$ for 100 epochs. It took approximately 10 hours to train our model on an NVIDIA RTX2070 GPU.
\begin{figure}[h!t]
    \centering
    \includegraphics[height=15cm,width=\columnwidth]{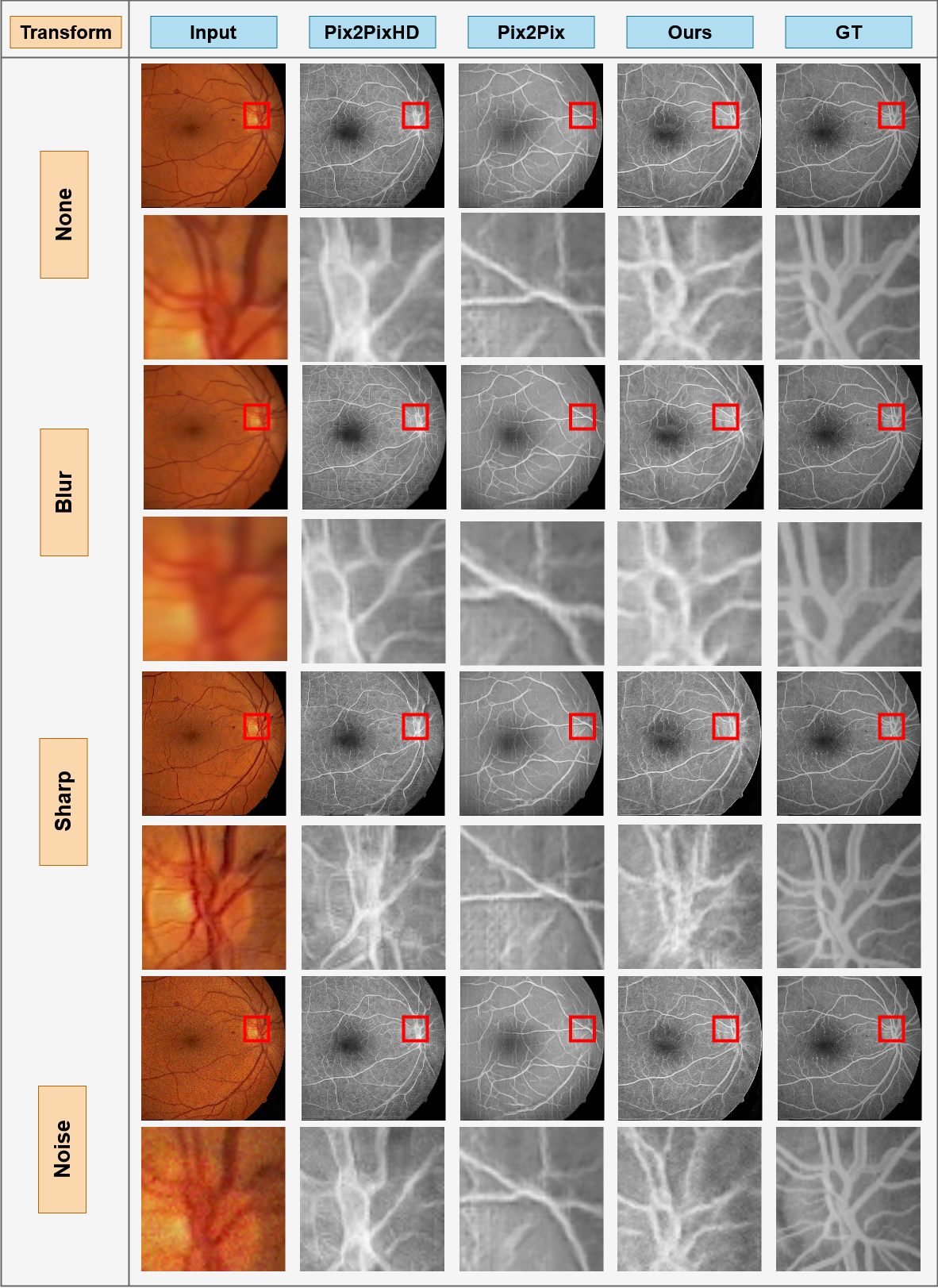}
    \caption{Angiogram generated from transformed Fundus images. The first row shows the original image, ground truth and generated angiograms from different architectures. The second row shows the closed-up version of a selected region. The red bounding box signifies that specific region. Each row pairs illustrated None, Blur, Sharp and Noise transformations.}
    \label{fig4}
\end{figure}
\subsection{Qualitative Evaluation}
\label{subsec:quant}
For evaluating the performance of the network, we took 14 images and cropped 4 sections from each quadrant of the image with a size of $512\times512$. We conducted two sets of experiments to evaluate both the network's robustness to global changes to the imaging modes and its ability to adapt to structural changes to the vascular patterns and structure of the eye. We used GNU Image Manipulation Program (GIMP) \cite{gimp2019gimp} for transforming and distorting images.
 
In the first set of experiments, three transformations were applied to the images: 1) blurring to represent out of focus funduscopy or fundus photography in the presence of severe cataracts, 2) sharpening to represent pupil dilation, and 3) noise to represent interference during photography. Good robustness is represented by the generated angiograms similarity to the real FA image since these transformation do not affect the vascular structure of the retina. A side by side comparison of different architecture's prediction is shown in Fig.~\ref{fig4}. As it can be observed from the image, the proposed architecture produces images very similar to the ground-truth (GT) under these global changes applied to the fundus image. 
\begin{figure}[t]
    \centering
    \includegraphics[width=\columnwidth]{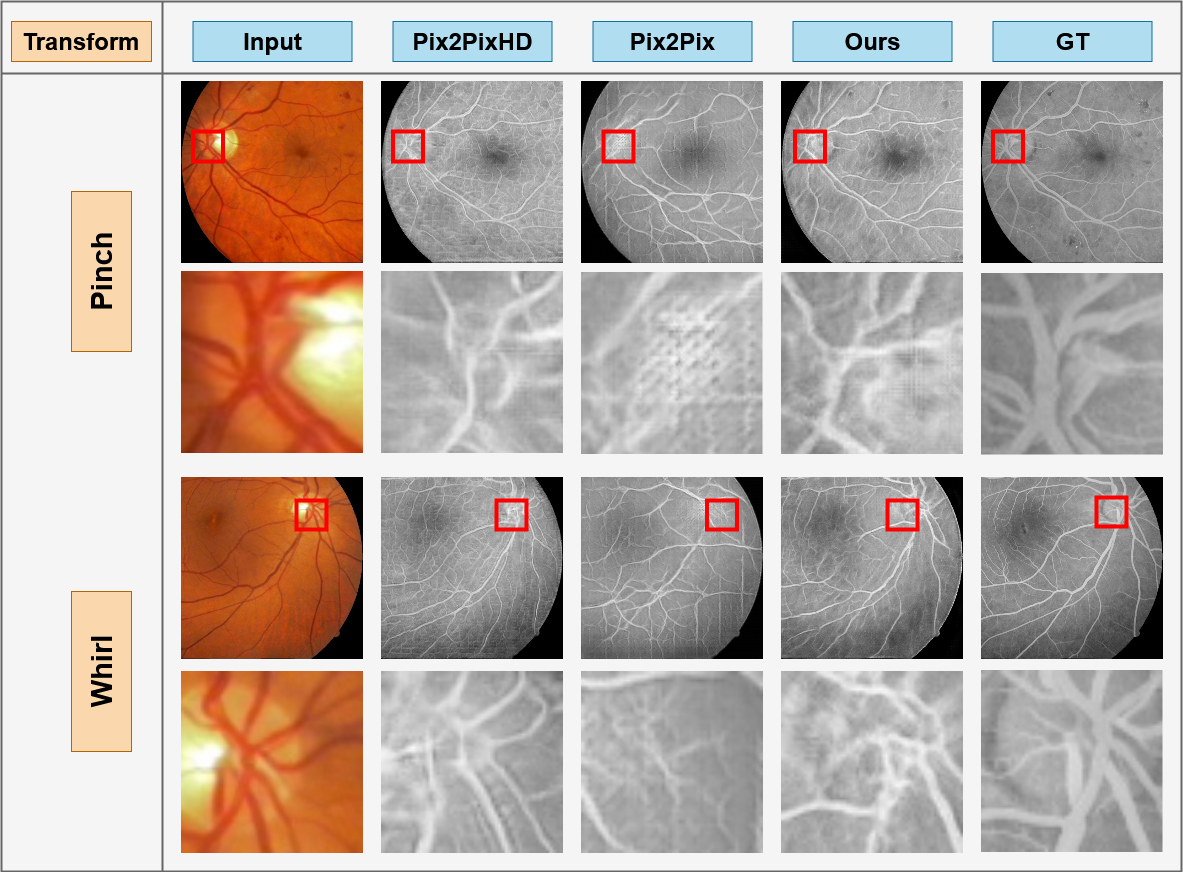}
    \caption{Angiogram generated from distorted Fundus images with biological markers. The first row shows the original image, ground truth and generated angiograms from different architectures. The second row shows the closed-up version of a selected region. The red bounding box signifies that specific region. Each row pairs illustrated Pinch and Whirl transformations.}
    \label{fig5}
\end{figure}

In the case of \textbf{blurred} fundus images, our model is less affected compared to other architectures, as seen in the second row of Fig.~\ref{fig4}-- structure of smaller veins are preserved better compared to Pix2Pix and Pix2PixHD.

In the case of \textbf{sharpened} images, the angiogram produced by Pix2Pix and Pix2PixHD show vein-like structures introduced in the back, which are not present in our prediction. These are seen in the third row of Fig.~\ref{fig4}.

In the case of \textbf{noisy} images, as seen in the last row of Fig.~\ref{fig4} our prediction is still unaffected with this pixel level alteration. However, both Pix2Pix and Pix2PixHD fails to generate thin and small vessel structures by failing to extract low level features.

In the second set of experiments we modified the vascular pattern of the retina and the fundus images. These structural changes are represented by two different types of distortions: 1) pinch, representing the flattening of the retina resulting in the pulled/pushed retinal structure, and 2) whirl, representing retina distortions caused by increased intra-ocular pressure (IOP). Good adaptation to structural changes in the retina is achieved if the generated angiograms are similar to the angiograms with changed vascular structure. The effects of Pinch and Whirl on predicted angiogram is illustrated in Fig.~\ref{fig5}. 

\textbf{Pinch} represents the globe flattening condition, manifesting vascular changes on the retina as a result of distortions of retinal subspace. This experiment shows the adaptability and reproduciblity of the proposed network to uncover the changes in vascular structure. From the first row in Fig.~\ref{fig5} it is evident that our model can effectively locates the retinal vessels compared to other proposed techniques.

\textbf{Whirl} represented changes in the IOP or vitreous changes in the eye that may result in twists in the vascular structure. Similar to pinch, the network's ability to adapt to this structural change can be measured if the generated FA image is similar to the real angiogram showing the changed vascular structure. As seen in the last row of Fig.~\ref{fig5} our network encodes the feature information vessel structures, and is much less affected this kind of distortion. The other architectures failed to generate micro vessel structure as it can be seen in Fig.~\ref{fig5}.

\subsection{Quantitative Evaluations}
\label{subsec:qual}

For quantitative evaluation, we also performed two experiments. In the first experiment we use the Fréchet inception distance (FID) \cite{heusel2017gans} that has been used to evaluate similar style-transfer GANs \cite{karras2019style}. We computed the FID scores for different architectures on the generated FA image and original angiogram, and those generated from the changed fundus images by the five global and structural changes --i.e., blurring, sharpening, noise, pinch, and whirl. The results are reported in Table.~\ref{table2}. It should be noted that, lower FID score means better results. 

\begin{table}[htb]
\caption{Fréchet inception distance (FID) for different architectures}
    \label{table2}
\centering
\begin{tabular}{l|c|c|c|c|c|c} 
\hline
\small Architecture & \small Orig. & \small Noise & \small Blur & \small Sharp & \small Whirl & \small Pinch \\
\hline
\small \textbf{Ours} & \small\textbf{ 30.3} & \small \textbf{41.5 }(11.2$\uparrow$) & \small \textbf{32.3} (2.0$\uparrow$) & \small \textbf{34.3} (4.0$\uparrow$) & \small \textbf{38.2} (7.9$\uparrow$) & \small \textbf{33.1} (2.8$\uparrow$) \\ 

\small Pix2PixHD \cite{wang2018high} & \small 42.8  &  \small 53.0 (10.2$\uparrow$)& \small 43.7 (1.1$\uparrow$) & \small 47.5 (4.7$\uparrow$) & \small 45.9 (3.1$\uparrow$) & \small 39.2 (3.6$\downarrow$) \\ 
\small Pix2Pix \cite{isola2017image} & \small 48.6  &  \small 46.8 (1.8 $\downarrow$)& \small 50.8 (2.2$\uparrow$)& \small 47.1 (1.5$\downarrow$) & \small 43.0 (5.6$\downarrow$)& \small 43.7 (4.9$\downarrow$) \\ 
\hline
\end{tabular}
\end{table}

From Table.~\ref{table2}, using the original fundus image, the FID of our network angiogrm is 30.3, while other techniques are at least 10 points worse, Pix2PixHD (42.8) and Pix2Pix (48.6). For the case of noisy images, the FID for Pix2Pix dropped slightly but increased for both Pix2PixHD and our technique. Notice that the FID for our technique is still better than both Pix2Pix and Pix2PixHD. For all other changes, the FID score of our technique increased slightly but still outperformed Pix2Pix and Pix2PixHD in both robustness and adaptation to the structural changes.

\begin{table}[htb]
\caption{Results of Qualitative with Undisclosed Portion of Fake/Real Experiment}
    \label{table3}
\centering
\begin{tabular}{l|c|c|c|c|c} 
\hline
&\multicolumn{2}{c|}{\small Results} & \multicolumn{2}{c|}{ \small Average} & \\\hline
& \small Correct & \small Incorrect & \small Missed & \small Found & \small Confusion\\\hline
\small Fake & \small 15\% & \small 85\% & \small \multirow{2}{*}{53\%} & \small \multirow{2}{*}{48\%} & \multirow{2}{*}{\small \textbf{52.5\%}} \\
\small Real & \small 80\% & \small 20\% & & & \\
\hline
\end{tabular}
\end{table}

In the next experiment we evaluate the quality of the generated angiograms by asking experts (e.g. ophthalmologists) to identify fake angiograms among a collection of 40 balanced (50\%, 50\%) and randomly mixed angiograms. For this experiment, the experts were not told how many of the images are real and how many are fake. The non-disclosed ratio of fake and real images was a significant design choice for this experiment, as it will allow us to evaluate three metrics: 1) incorrectly labeled fake images representing how real the generated images look, 2) correctly labeled real images representing how accurate the experts recognized angiogram salient features, and 3) the confusion metric representing how effective the overall performance of our proposed method was in confusing the expert in the overall experiment. The results are shown in Table~\ref{table3}.

As it can be seen from Table~\ref{table3}, experts assigned 85\% of the fake angiogams as real. This result shows that experts had difficulty in identifying fake images, while they easily identified real angiograms with 80\% accuracy. Overall, the experts misclassified 53\% of all images. This resulted in a confusion factor of 52.5\%. This is significant, as the confusion factor of 50\% is the best achievable result.

\section{Conclusion}

In this paper, we introduced Fundus2Angio, a novel conditional generative architecture that capable of generating angiograms from retinal fundus images. We further demonstrated its robustness, adaptability, and reproducibility by synthesizing high quality angiograms from transformed and distorted fundus images. Additionally, we illustrated how changes in biological markers do not affect the adaptability and reproducibility of  synthesizing angiograms by using our technique. This ensures that the proposed architecture effectively preserves known biological markers (e.g. vascular patterns and structures). As a result, the proposed network can be effectively utilized to produce accurate FA images for the same patient from his or her fundus images over time. This allows for a better control on patient's disease progression monitoring or to help uncover newly developed diseases or conditions. One future direction to this work is to improve upon this work to incorporate retinal vessel segmentation and exudate localization.

\bibliographystyle{splncs04}
\bibliography{references}

\begin{thebibliography}{10}
\providecommand{\url}[1]{\texttt{#1}}
\providecommand{\urlprefix}{URL }
\providecommand{\doi}[1]{https://doi.org/#1}

\bibitem{brockow2014hypersensitivity}
Brockow, K., S{\'a}nchez-Borges, M.: Hypersensitivity to contrast media and
  dyes. Immunology and Allergy Clinics  \textbf{34}(3),  547--564 (2014)

\bibitem{brown2003recognising}
Brown, M., Lowe, D.G., et~al.: Recognising panoramas. In: ICCV. vol.~3, p.~1218
  (2003)

\bibitem{chen2017photographic}
Chen, Q., Koltun, V.: Photographic image synthesis with cascaded refinement
  networks. In: Proceedings of the IEEE international conference on computer
  vision. pp. 1511--1520 (2017)

\bibitem{chen2018sketchygan}
Chen, W., Hays, J.: Sketchygan: Towards diverse and realistic sketch to image
  synthesis. In: Proceedings of the IEEE Conference on Computer Vision and
  Pattern Recognition. pp. 9416--9425 (2018)

\bibitem{chollet2017xception}
Chollet, F.: Xception: Deep learning with depthwise separable convolutions. In:
  Proceedings of the IEEE conference on computer vision and pattern
  recognition. pp. 1251--1258 (2017)

\bibitem{hajeb2012diabetic}
Hajeb Mohammad~Alipour, S., Rabbani, H., Akhlaghi, M.R.: Diabetic retinopathy
  grading by digital curvelet transform. Computational and mathematical methods
  in medicine  \textbf{2012} (2012)

\bibitem{he2016identity}
He, K., Zhang, X., Ren, S., Sun, J.: Identity mappings in deep residual
  networks. In: European conference on computer vision. pp. 630--645. Springer
  (2016)

\bibitem{heusel2017gans}
Heusel, M., Ramsauer, H., Unterthiner, T., Nessler, B., Hochreiter, S.: Gans
  trained by a two time-scale update rule converge to a local nash equilibrium.
  In: Advances in neural information processing systems. pp. 6626--6637 (2017)

\bibitem{huang2017stacked}
Huang, X., Li, Y., Poursaeed, O., Hopcroft, J., Belongie, S.: Stacked
  generative adversarial networks. In: Proceedings of the IEEE conference on
  computer vision and pattern recognition. pp. 5077--5086 (2017)

\bibitem{isola2017image}
Isola, P., Zhu, J.Y., Zhou, T., Efros, A.A.: Image-to-image translation with
  conditional adversarial networks. In: Proceedings of the IEEE conference on
  computer vision and pattern recognition. pp. 1125--1134 (2017)

\bibitem{johnson2016perceptual}
Johnson, J., Alahi, A., Fei-Fei, L.: Perceptual losses for real-time style
  transfer and super-resolution. In: European conference on computer vision.
  pp. 694--711. Springer (2016)

\bibitem{opticnet19}
{Kamran}, S.A., {Saha}, S., {Sabbir}, A.S., {Tavakkoli}, A.: Optic-net: A novel
  convolutional neural network for diagnosis of retinal diseases from optical
  tomography images. In: 2019 18th IEEE International Conference On Machine
  Learning And Applications (ICMLA). pp. 964--971 (2019)

\bibitem{karras2019style}
Karras, T., Laine, S., Aila, T.: A style-based generator architecture for
  generative adversarial networks. In: Proceedings of the IEEE Conference on
  Computer Vision and Pattern Recognition. pp. 4401--4410 (2019)

\bibitem{li2016precomputed}
Li, C., Wand, M.: Precomputed real-time texture synthesis with markovian
  generative adversarial networks. In: European conference on computer vision.
  pp. 702--716. Springer (2016)

\bibitem{lira2007adverse}
Lira, R.P.C., Oliveira, C.L.d.A., Marques, M.V.R.B., Silva, A.R., Pessoa,
  C.d.C.: Adverse reactions of fluorescein angiography: a prospective study.
  Arquivos brasileiros de oftalmologia  \textbf{70}(4),  615--618 (2007)

\bibitem{mandava2004fluorescein}
Mandava, N., Reichel, E., Guyer, D., et~al.: Fluorescein and icg angiography.
  St Louis: Mosby  \textbf{106},  800--808 (2004)

\bibitem{mao2017least}
Mao, X., Li, Q., Xie, H., Lau, R.Y., Wang, Z., Paul~Smolley, S.: Least squares
  generative adversarial networks. In: Proceedings of the IEEE International
  Conference on Computer Vision. pp. 2794--2802 (2017)

\bibitem{mary2016retinal}
Mary, V.S., Rajsingh, E.B., Naik, G.R.: Retinal fundus image analysis for
  diagnosis of glaucoma: a comprehensive survey. IEEE Access  (2016)

\bibitem{pathak2016context}
Pathak, D., Krahenbuhl, P., Donahue, J., Darrell, T., Efros, A.A.: Context
  encoders: Feature learning by inpainting. In: Proceedings of the IEEE
  conference on computer vision and pattern recognition. pp. 2536--2544 (2016)

\bibitem{poplin2018prediction}
Poplin, R., Varadarajan, A.V., Blumer, K., Liu, Y., McConnell, M.V., Corrado,
  G.S., Peng, L., Webster, D.R.: Prediction of cardiovascular risk factors from
  retinal fundus photographs via deep learning. Nature Biomedical Engineering
  \textbf{2}(3), ~158 (2018)

\bibitem{shaham2019singan}
Shaham, T.R., Dekel, T., Michaeli, T.: Singan: Learning a generative model from
  a single natural image. In: Proceedings of the IEEE International Conference
  on Computer Vision. pp. 4570--4580 (2019)

\bibitem{gimp2019gimp}
Team, G., et~al.: GIMP: GNU Image Manipulation Program. GIMP Team. (2019)

\bibitem{wang2018high}
Wang, T.C., Liu, M.Y., Zhu, J.Y., Tao, A., Kautz, J., Catanzaro, B.:
  High-resolution image synthesis and semantic manipulation with conditional
  gans. In: Proceedings of the IEEE conference on computer vision and pattern
  recognition. pp. 8798--8807 (2018)

\bibitem{yannuzzi1986fluorescein}
Yannuzzi, L.A., Rohrer, K.T., Tindel, L.J., Sobel, R.S., Costanza, M.A.,
  Shields, W., Zang, E.: Fluorescein angiography complication survey.
  Ophthalmology  \textbf{93}(5),  611--617 (1986)

\bibitem{zhu2016generative}
Zhu, J.Y., Kr{\"a}henb{\"u}hl, P., Shechtman, E., Efros, A.A.: Generative
  visual manipulation on the natural image manifold. In: European Conference on
  Computer Vision. pp. 597--613. Springer (2016)

\bibitem{zhu2017unpaired}
Zhu, J.Y., Park, T., Isola, P., Efros, A.A.: Unpaired image-to-image
  translation using cycle-consistent adversarial networks. In: Proceedings of
  the IEEE international conference on computer vision. pp. 2223--2232 (2017)

\end{thebibliography}

\end{document}